\RequirePackage{fix-cm}
\documentclass[twocolumn,epjc3]{svjour3}  
\RequirePackage{graphicx}
\usepackage{amsmath}
\usepackage{bm}% bold math
\journalname{Eur. Phys. J. E}

\begin{document}

\title{Memory effect of external oscillation on residual stress in a paste}


\author{Jumpei Morita\thanksref{e1,addr1}
        \and
	Michio Otsuki\thanksref{e2,addr2} %.
}

\thankstext{e1}{e-mail: j.morita0612@gmail.com (Current affiliation is Neturen Co., Ltd.)}
\thankstext{e2}{e-mail: otsuki@me.es.osaka-u.ac.jp}


\institute{Department of Physics and Material Science, Shimane University, Matsue, Shimane 690-8504, Japan.\label{addr1}
           \and
           Graduate School of Engineering Science, Osaka University, Toyonaka, Osaka 560-8531, Japan.\label{addr2}
}

\date{Received: date / Accepted: date}

\abstractdc{
  We numerically investigate the stress distribution of a paste when an external oscillation is applied.
  The paste memorizes the oscillation through plastic deformation.
  Due to the plastic deformation, the residual stress remains after the oscillation, where
 the residual stress distribution depends on the number of cycles in the oscillation. 
  As this number increases, the symmetry of the stress distribution is enhanced, which is consistent with the crack patterns observed in the experiments using a drying paste.
}

\maketitle

\section{Introduction}

When a paste containing powder and water is dried, cracks are formed \cite{Goehring}. 
Crack patterns on the surface of the paste are usually random and isotropic in shallow containers \cite{Groisman}.
However, recent experiments have revealed that the crack patterns become anisotropic on application of external fields \cite{Nakahara05,Nakahara06a,Nakahara06b,Matsuo,Nakayama,Akiba,Mal,Khatun,Pauchard,Ngo,Lama}.
In particular, Nakahara and Matsuo reported that subjecting the container to horizontal oscillation before desiccation imprints a ``memory'' in the paste.
This memory induces lamellar crack patterns perpendicular to the direction of the oscillation after several days \cite{Nakahara05}.

They also discovered that such crack patterns appear when the stress induced by the oscillation exceeds the yield stress of the paste \cite{Nakahara05}.
This experimental result indicates that the plastic deformation caused by the oscillation is related to the memory effect. 
The visualization of the plastic deformation in the drying paste showing the memory effect supports this conjecture \cite{Goehring}.

The cracks are formed to release the tension caused by the desiccation when a fracture criterion is satisfied \cite{Beuth,Kitsunezaki,Singh,Man}.
Therefore, the memory effect can be explained if plastic deformation affects
one of the factors in crack formation, namely the tension or
the fracture criterion.
In this regard, recent experiments to measure stresses in drying paste with memory effect \cite{Kitsunezaki16,Kitsunezaki17} deserve special attention.
According to these experiments, the fracture criterion is unchanged, but the external oscillation induces the residual stress in the direction of the oscillation, which increases due to the desiccation and leads to the perpendicular crack pattern.
However, the mechanism by which plastic deformation causes residual stress is not clear in the experiments.

The stress-measuring experiments \cite{Kitsunezaki16,Kitsunezaki17} had been motivated by the theoretical conjecture that anisotropy in residual stresses caused by plastic deformation is responsible for the memory effect \cite{Otsuki,Ooshida08,Ooshida09}.
However, theoretical predictions in the previous studies seem to be incoherent. 
On the one hand, a quasi-linear analysis of an elastoplastic model based on the infinitesimal strain theory was performed in Ref. \cite{Otsuki}.
In this analysis, a non-uniform plastic deformation remains after the external oscillation owing to the boundary condition at the walls of the container.
The plastic deformation causes an asymmetric stress distribution in the direction of the oscillation, and the tension increases in some parts of the paste. 
The desiccation after the oscillation enhances the asymmetric stress, which creates cracks perpendicular to the direction of the external oscillation.
Therefore, in this analysis, the crack patterns are asymmetric.
On the other hand, in Ref. \cite{Ooshida08,Ooshida09}, a nonlinear analysis based on finite strain theory is performed. 
This analysis shows that a uniform plastic deformation under periodic boundary conditions increases the tension over the entire area of the paste. 
This uniform increase of the tension leads to uniform and symmetric crack patterns along the direction of the oscillation.

However, the seemingly incoherent predictions may represent two different limiting cases, as is suggested by a recent experiment \cite{Nakahara18}, where the symmetric and asymmetric crack patterns occur depending on the number of cycles in the external oscillation.
For a few cycles, the asymmetric crack pattern predicted in the quasi-linear analysis was formed. 
For a sufficiently large number of cycles, the symmetric pattern predicted in the nonlinear analysis was formed.
However, the present theories cannot explain the dependence of the crack pattern on the number of cycles.
Therefore, a unified theoretical model is required.

In this study, we formulate such a unified model, with which we numerically examine the residual stress in a paste after external oscillation.
In Sec. \ref{sec:kinematics}, we review a Lagrangian description of kinematics. 
In Sec.  \ref{sec:motion}, we derive the time evolution equations for a paste. 
The memory effect of the external oscillation is numerically investigated in Sec. \ref{sec:memory}.
The setup of our simulation is shown in Sec. \ref{sec:setup}. 
We demonstrate the relation between the plastic deformation and the residual stress in Sec. \ref{sec:previous}. 
In Secs. \ref{sec:half} and \ref{sec:two}, we present the residual stress after the oscillation for different numbers of cycles, respectively. 
The dependence of the stress distribution on the number of cycles is examined in Sec. \ref{sec:number}.
We discuss our results and conclude our paper in Sec. \ref{sec:discussion}. 
In \ref{sec:CGT}, the left Cauchy--Green tensor in our setup is derived.
The stretching tensor is obtained in \ref{sec:ST}.
The constitutive relation of the paste is derived in \ref{sec:stress}.
In \ref{sec:steady}, we discuss the shear stress under steady shear in our model.

\section{Kinematics}
\label{sec:kinematics}

The configuration of a 3-dimensional paste is represented by the current coordinates $\bm{r}$ mapped from the reference coordinates $\bm{X}$ as
\begin{eqnarray}
\label{map 1}
( \bm{X} , t ) \mapsto
\bm{r} ( \bm{X} , t ) = 
\left[ 
\begin{array}{ccc}
x ( X,Y,Z, t ) \\
y ( X,Y,Z, t ) \\
z ( X,Y,Z, t ) \\
\end{array} 
\right]_{\rm C},
\end{eqnarray}
where $[\cdot]_{\rm C}$ denotes the representation in terms of Cartesian components.
The velocity $\bm{v}$ is given by
\begin{eqnarray}
\label{Velocity}
\bm{v}( \bm{X} , t ) = \partial_t \bm{r}( \bm{X} , t ),
\end{eqnarray}
and the acceleration is given by $\partial_t^2 \bm{r}$.
Here, $\partial_t$ stands for the time derivative in the Lagrangian description,
and the independent variables of the paste are $\bm{X}$ and $t$.
Therefore, $\partial_t$ represents the derivative with respect to $t$, with $\bm{X}$ being constant.

The line element $d\bm{r}$ between two points labeled as $\bm{X}$ and $\bm{X} + d\bm{X}$ with an infinitesimal vector $d\bm{X}$ is given by
\begin{eqnarray}
d\bm{r} = ( \partial _{i} \bm{r} ) d X^{i}, 
  \label{dr}
\end{eqnarray}
where $\partial _{i}  = \partial / \partial X^i$, and we use the Einstein notation for the summation of repeated indices.
The square of the distance between the points in the current configuration
is 
\begin{equation}
(ds)^2
= | ( \partial _{i} \bm{r} ) d X^{i} |^{2} 
= g_{ij} d X^{i} d X^{j},
\label{Euclidean metric 1}
\end{equation}
where $g_{ij}$ is a component of the Euclidean metric tensor $\bm{g}$ given by
\begin{equation}
\label{Euclidean metric 2}
g_{ij} = (\partial _{i} \bm{r}) \cdot (\partial _{j} \bm{r}).
\end{equation}

Assuming the existence of a local stress-free ``natural state'', the line element between the two points in this state is denoted as $d\bm{r}^\natural$ \cite{Ooshida08}. 
The square of the distance between the points in the ``natural state'' is given by
\begin{equation}
\label{Natural metric 1}
( ds^{\natural} )^{2} = |d\bm{r}^\natural|^2
= g_{ij}^{\natural} d X^{i} d X^{j},
\end{equation}
where $g_{ij}^{\natural}$ is the natural metric tensor satisfying $g_{ij}^{\natural} = g_{ji}^{\natural}$.
The change in the natural metric tensor represents the plastic deformation.

The elastic deformation is represented by the transformation from $d\bm{r}^\natural$ to $d\bm{r}$:
\begin{equation}
\label{Deformation gradient}
d\bm{r} = \mathbf{F} d\bm{r}^\natural,
\end{equation}
where $\mathbf{F}$ is the deformation gradient tensor.
The left Cauchy--Green Tensor is defined as \cite{Romano,Truesdell}
\begin{equation}
\mathbf{B}  = \mathbf{F} \mathbf{F}^T.
  \label{B}
\end{equation}
$\mathbf{B}$ is given by
\begin{equation}
\mathbf{B}  = g^{ij}_{\natural} ( \partial _{i} \bm{r} ) \otimes ( \partial _{j} \bm{r} ),
\label{LCG4}
\end{equation}
where $(g^{ij}_{\natural} )$ denotes the inverse of the component matrix of the metric tensor $(g_{ij}^{\natural} )$ satisfying
\begin{equation}
g^{\natural}_{ij} g_{\natural}^{jk} = \delta_i^k.
\label{Natural metric 2}
\end{equation}
See \ref{sec:CGT} for the derivation of Eq. \eqref{LCG4}.

The difference in velocities  between two points labeled as $\bm{X}$ and $\bm{X} + d\bm{X}$ is given by
\begin{equation}
  d\bm{v} = \left ( \partial_i  \bm{v} \right )dX^i.
\label{dv}
\end{equation}
The time rate of the deformation is described by the velocity gradient tensor $\mathbf{W}$, which satisfies
\cite{Romano,Truesdell}
\begin{equation}
\label{Velocity gradient}
d\bm{v} = \mathbf{W} d\bm{r}.
\end{equation}
The stretching tensor is defined as
\begin{equation}
\mathbf{D} = \frac{1}{2} \left ( \mathbf{W} +  \mathbf{W}^T \right ).
  \label{stretching}
\end{equation}
As shown in \ref{sec:ST}, the stretching tensor is represented as
\begin{equation}
 \mathbf{D} = - \frac{1}{2} \left (\partial_t g^{ij} \right )
( \partial _{i} \bm{r} ) \otimes ( \partial _{j} \bm{r} ),
\label{strain rate}
\end{equation}
where $(g^{ij})$ is the inverse of the component matrix of the metric tensor $(g_{ij})$ satisfying
\begin{equation}
g_{ij} g^{jk} = \delta_i^k.
\label{Orthogonal}
\end{equation}

\section{Equation of motion and constitutive relation}
\label{sec:motion}

The equation of motion for $\bm{r}(\bm{X},t)$ is given by 
\begin{equation}
\label{Cauchys momentum equation 1}
\rho \partial^{2}_{t} \bm{r} = {\rm div} \boldsymbol{\sigma} + \bm{f}_{\rm e},
\end{equation}
where $\rho$, $\boldsymbol{\sigma}$, and $\bm{f}_{\rm e}$ are the density in the current configuration, the Cauchy stress tensor, and the body force, respectively.
The Cauchy stress tensor consists of an elastoplastic part $\boldsymbol{\sigma}^{\rm (EP)}$ and a viscous part $\boldsymbol{\sigma}^{\rm (V)}$ as 
\begin{equation}
  \boldsymbol{\sigma} =  \boldsymbol{\sigma}^{\rm (EP)} +  \boldsymbol{\sigma}^{\rm (V)}.
\end{equation}

The elastoplastic part is given by the constitutive relation with the strain energy per unit volume in the reference configuration $\Sigma_{\rm R}$ as
\begin{equation}
\label{Constitutive}
\boldsymbol{\sigma}^{\rm (EP)}  = \frac{1}{J } 
\frac{\partial \Sigma_{\rm R}}{\partial \tilde{\mathbf{F}}}\  \tilde{\mathbf{F}}^T,
\end{equation}
where $\tilde{\mathbf{F}} = \partial \bm{r} / \partial \bm{X}$ is the ``apparent'' deformation gradient tensor characterizing the transformation from the reference configuration to the current configuration \cite{Beatty}.
Here, $J = \det \tilde{\mathbf{F}}  = \sqrt{\det \bm{g}}$ is the Jacobian.
The strain energy per unit volume in the current coordinates $\Sigma$ is related to $\Sigma_{\rm R}$ as
\begin{eqnarray}
\label{Strain Energies}
  \Sigma_{\rm R} & = & J  \Sigma.
\end{eqnarray}
In this study, we apply the Hadamard strain energy \cite{destrade}:
\begin{eqnarray}
\label{Hadamard 1}
  \Sigma & = & \left \{ \mu ( I_{1} - 3 ) + \Psi ( I_{3} ) \right \}/2, \\
\label{Hadamard 2}
  \Psi ( I_{3} ) & = & ( \lambda + \mu ) ( I_{3} - 1 ) - 2 ( \lambda + 2 \mu ) ( \sqrt{ I_{3} } - 1 ),
\end{eqnarray}
where $I_{1} = \rm{tr} \mathbf{B}$ and $I_{3} = \rm{det} \mathbf{B} $ are the rotational invariants of the left Cauchy--Green strain tensor $\mathbf{B}$ with the Lam\'{e} constants $\mu$ and $\lambda$.
Note that $I_1$ and $I_3$ depend on $\tilde{\mathbf{F}}$, as shown in \ref{sec:stress}.
In Ref. \cite{Ooshida08}, the incompressible neo-Hookean model is adopted for the strain energy, but we assume Hadamard strain energy with small compressibility (i.e. large but finite $\lambda$) to avoid difficulties in numerical simulations. 
Note that the neo-Hookean strain energy is the incompressible limit of the Hadamard strain energy.
Substituting Eq. \eqref{Strain Energies} with Eqs. \eqref{Hadamard 1} and \eqref{Hadamard 2} into Eq. \eqref{Constitutive}, we obtain
\begin{equation}
\label{stress 2}
\boldsymbol{\sigma}^{\rm (EP)} = \{\Sigma + I_{3} \Psi^{\prime} ( I_{3} ) + \mu \} \mathbf{I} + \mu \left( \mathbf{B} - \mathbf{I} \right).
\end{equation}
Here, $\mathbf{I}$ is the unit tensor.
See \ref{sec:stress} for the derivation of Eq. \eqref{stress 2}.
For $\boldsymbol{\sigma}^{\rm (V)}$, we adopt the linear viscous stress tensor as
\begin{equation}
\label{viscosity 1}
\boldsymbol{\sigma}^{\rm (V)} =
2 \eta \mathbf{D}
\end{equation}
with the viscosity $\eta$ \cite{Romano}.

Following Ref. \cite{Ooshida08,Ooshida09}, the plastic deformation is described by the temporal evolution of the natural metric $g_{\natural}^{ij}$ as
\begin{equation}
\label{Maxwell 5}
( 1 + \tau \partial _{t} ) g_{\natural}^{ij} = K g^{ij}.
\end{equation}
Here, $K$ is given by 
\begin{equation}
\label{K2}
K = \frac{3}{g^{\natural}_{ij} g^{ij}}= \frac{3}{I_1}
\end{equation}
to satisfy the incompressibility condition $J^\natural =  \sqrt{\det \bm{g}^\natural}  = 1$ in the natural state.
It should be noted that we assume effective incompressibility for $\bm{g}$ by adopting Poisson's ratio $\nu \equiv \lambda/(\lambda+\mu)/2 \simeq 1/2$ in our numerical simulations.
The inverse of the relaxation time $\tau$ is given by
\begin{equation}
\label{Bingham 3}
  \tau^{-1}( \sigma_{\rm e} ) = \frac{1}{\tau_0}{\rm max} \left( 0 , 1 - \frac{ \sigma _{\rm Y} }{ \sigma_{\rm e} } \right)
\end{equation}
with characteristic time $\tau_0$ and tensile yield strength $\sigma _{\rm Y}$.
The equivalent tensile stress $\sigma_{\rm e}$ is given by
\begin{eqnarray}
\sigma_{\rm e}^2 & = & \frac{1}{2} (\sigma_{xx}^{\rm (EP)} - \sigma_{yy}^{\rm (EP)} )^2 \nonumber \\
  & & + \frac{1}{2} (\sigma_{yy}^{\rm (EP)} - \sigma_{zz}^{\rm (EP)} )^2 \nonumber \\
  & & +\frac{1}{2} (\sigma_{zz}^{\rm (EP)} - \sigma_{xx}^{\rm (EP)} )^2, 
\nonumber \\
  & & +3 \left \{ \left(\sigma_{xz}^{\rm (EP)}\right)^2+\left(\sigma_{yz}^{\rm (EP)}\right)^2+\left(\sigma_{zx}^{\rm (EP)}\right)^2 \right \}.
\label{von Mises}
\end{eqnarray}
Equations \eqref{Maxwell 5} and \eqref{Bingham 3} indicate that the plastic deformation associated with the change of the natural metric $g_{\natural}^{ij}$ occurs when the von Mises yield criterion $\sigma_{\rm e} = \sigma_{\rm Y}$ is satisfied \cite{Jones}.
See \ref{sec:steady} for the flow curve of our model under steady shear.

\section{Memory of external oscillation}
\label{sec:memory}

In this section, we numerically investigate the memory effect of the external oscillation.
In Sec. \ref{sec:setup}, we explain our setup.
In Sec. \ref{sec:previous}, we demonstrate how plastic deformation affects the residual stress.
The residual stress for different numbers of cycles is shown in Secs. \ref{sec:half} and \ref{sec:two}.
In Sec. \ref{sec:discussion}, we discuss the dependence of the symmetry of the stress distribution on the number of cycles.

\subsection{Setup}
\label{sec:setup}

\begin{figure}[htbp]
  \centering
  \includegraphics[width=1.0\linewidth]{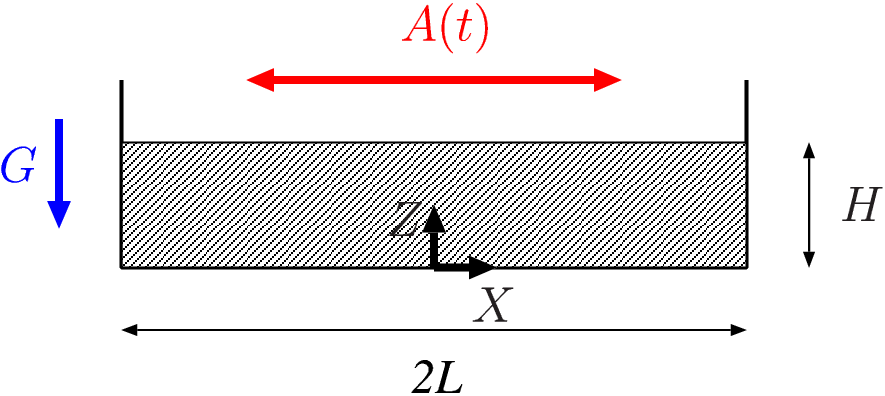}
  \caption{Schematic illustration of a paste in a container.}
  \label{fig:2d}
\end{figure}

Let us consider a paste of thickness $H$ and width $2L$ in a container, as shown in Fig. \ref{fig:2d}.
The center of the container is at $X=0$, and the bottom is at $Z=0$.
We assume plane strain deformation so that the current configuration is given by
\begin{eqnarray}
\label{map2}
\bm{r} ( X , Z , t ) = 
\left[ 
\begin{array}{ccc}
X + u ( X , Z , t ) \\
Y \\
Z + w ( X , Z , t ) \\
\end{array} 
\right]_{\rm C}
\end{eqnarray}
with displacements $u$ and $w$ in the $X$ and $Z$ directions, respectively.
The body force is given by
\begin{eqnarray}
\bm{f}_{\rm e} = 
- \rho
\left[ 
\begin{array}{c}
A(t) \\
0 \\
G \\
\end{array} 
\right]_{\rm C},
\end{eqnarray}
where $G$ is the gravitational acceleration and $A(t)$ is the horizontal acceleration due to the external oscillation of the container.

Assuming that the incompressibility of the natural state ($\det \mathbf{g}^{\natural} = 1$), $\mathbf{g}^{\natural}$ can be expressed by two parameters, and we set
\begin{eqnarray}
\label{Natural metric3}
\mathbf{g}^{\natural} = 
\left[ 
\begin{array}{ccc}
  e^{ - \alpha }& 0 &  \beta \\
0 & 1 & 0 \\
 \beta & 0 & e^{ \alpha } ( 1 + \beta^{2} ) \\
\end{array} 
\right]
\end{eqnarray}
and
\begin{eqnarray}
\label{Natural metric4}
\mathbf{g}_{\natural} = 
\left[ 
\begin{array}{ccc}
  e^{ \alpha } ( 1 + \beta^{2} ) & 0 &  -\beta \\
0 & 1 & 0 \\
 - \beta & 0 & e^{ - \alpha } \\
\end{array} 
\right],
\end{eqnarray}
where $\beta$ represents the plastic shear strain, whereas the plastic normal strain is characterized by $\alpha$. $e^{-\alpha}$ represents a normal component of the natural metric tensor in the direction of the external oscillation.

The initial configuration is given by $u = w = \alpha = \beta= 0$ at $t=-T_{\rm I}$ with $g_{ij}^\natural = \delta_{ij}$.
For $-T_{\rm I} \le t < 0$, we set $A(t)= 0$ and relax the configuration under gravitational acceleration.
We apply the external oscillation as
\begin{eqnarray}
A(t)= -A_x \cos \frac{2 \pi t }{T}
\end{eqnarray}
for $0 \le t < N_{\rm c} T$ with the maximum acceleration $A_x$, the period $T$, and the number of cycles as $N_{\rm c}$.
After $N_{\rm c}$ cycles of the oscillation, we set $A(t) = 0$ and relax the system until $t = T_{\rm w}$.
The profiles of the stress and plastic deformation shown below are obtained at $t = T_{\rm w}$.

We assume the following no-slip boundary conditions at the bottom of the container:
\begin{eqnarray}
u|_{Z = 0} =  w|_{Z = 0} = 0,
\end{eqnarray}
while the stress applied to the free surface at the top of the paste is given by
\begin{eqnarray}
\sigma_{xz}|_{Z = H}=0, \ \ \sigma_{zz}|_{Z = H}=-p_0,
\end{eqnarray}
where $p_0$ is the atmospheric pressure.
For the lateral walls, we assume that the paste does not leave the wall as
\begin{eqnarray}
u|_{X = \pm L}= 0,
\end{eqnarray}
but it freely slides on the wall as
\begin{eqnarray}
\sigma_{zx}|_{X = \pm L} = 0.
\end{eqnarray}

In this study, we use the unit mass, length, and time as $m = \rho H^3$, $l = H$, and $\tau = \sqrt{H/G}$, respectively.
The parameter values are set as $L/H=10$, $A_x/G = 6.0 \times 10^{-2}$, $T/\tau = 31.25$, $\nu = \lambda/(\lambda+\mu)/2= 0.4999$, $\tau_0/\tau = 0.94$, $\mu /(\rho H G) =7.3 \times 10^{-2}$, $\sigma_{\rm Y}/(\rho H G) = 4.7 \times 10^{-2}$, $\eta/(\rho  \sqrt{GH^3})  = 0.14$, $T_I/\tau=100 $, and $T_w/\tau=300$ based on experiments \cite{Nakahara05}, except for $\mu$ and $\eta$.
It should be noted that $\tau_0$ is estimated from the flow curve under steady shear, which is shown in \ref{sec:steady}.
The atmospheric pressure $p_0$ is set to $0$ because the numerical results shown below do not depend on the value of $p_0$.
We have checked that our numerical results do not depend on the value of Poisson's ratio $\nu \equiv \lambda/(\lambda+\mu)/2$ for $\nu\ge 0.4999$.
We adopt the finite-difference method with the time interval as $\Delta t/\sqrt{H/G} =5.0 \times 10^{-5}$ and the spatial mesh size as $\Delta x/(2L) = \Delta z/(2L) = 0.025$.

\subsection{Effect of plastic deformation on residual stress}
\label{sec:previous}

In this section, we discuss the normal component of the deviatoric stress
\begin{eqnarray}
s_{xx} = \sigma_{xx} - (\sigma_{xx}+ \sigma_{yy}+ \sigma_{zz})/3
\end{eqnarray}
in the direction of the oscillation because its increase is essential for the formation of cracks.
In the linear approximation of $\partial_i u$, $\alpha$, and $\beta$ while ignoring $\partial_i w$, $s_{xx}$ is given by
\begin{eqnarray}
  s_{xx} = s_{xx}^{\rm (L)} + s_{xx}^{\rm (N)}
  \label{eq:s}
\end{eqnarray}
with 
\begin{eqnarray}
s_{xx}^{\rm (L)} = 2\mu u_X
  \label{eq:sL}
\end{eqnarray}
and
\begin{eqnarray}
s_{xx}^{\rm (N)} = 2\mu \alpha.
  \label{eq:sN}
\end{eqnarray}
Here, we abbreviate $\partial u / \partial X$ as $u_X$.
Finite $u_X$ due to the plastic deformation leads to the change of $s_{xx}$ through Eqs. \eqref{eq:s} and \eqref{eq:sL}, which corresponds to the mechanism of the memory effect assumed in the quasi-linear analysis \cite{Otsuki}. 
The plastic deformation characterized by $\alpha$ induces the increase of $s_{xx}$ through Eqs. \eqref{eq:s} and \eqref{eq:sN}, which is consistent with the scenario proposed in the nonlinear analysis \cite{Ooshida08,Ooshida09}.

\subsection{Residual stress for $N_c = 1/2$}
\label{sec:half}

%residual stress

\begin{figure}[htbp]
	\includegraphics[width=1.0\linewidth]{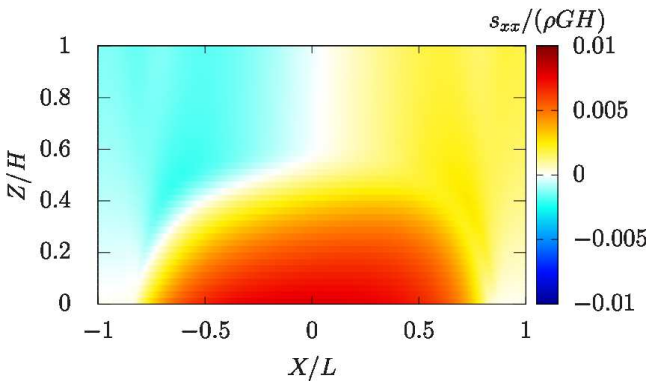}
    \caption{$s_{xx}$ as a function of $(X,Z)$ after the external oscillation with $N_c = 1/2$.}
    \label{fig:s_1/2}
\end{figure}

In Fig. \ref{fig:s_1/2}, we plot $s_{xx}$ as a function of $(X,Z)$ after the oscillation with $N_c = 1/2$ (i.e. half a cycle).
Here, $(X,Z)$ indicates the initial position.
During the oscillation, the equivalent tensile stress $\sigma_{\rm e}$ exceeds the tensile yield strength $\sigma_{\rm Y}$, and plastic deformation occurs at the bottom of the paste.
The residual stress remains as a memory effect due to the plastic deformation.
The residual stress distribution is almost symmetric with respect to the inversion of $X$ (i.e., $X \to -X$) and positive near the bottom, while it is asymmetric near the surface of the paste.
The asymmetric pattern near the surface is consistent with the stress distribution predicted in the quasi-linear analysis \cite{Otsuki}, which induces the asymmetric crack patterns observed in experiments with a few cycles of oscillation \cite{Nakahara18}.

\begin{figure}[htbp]
	\includegraphics[width=1.0\linewidth]{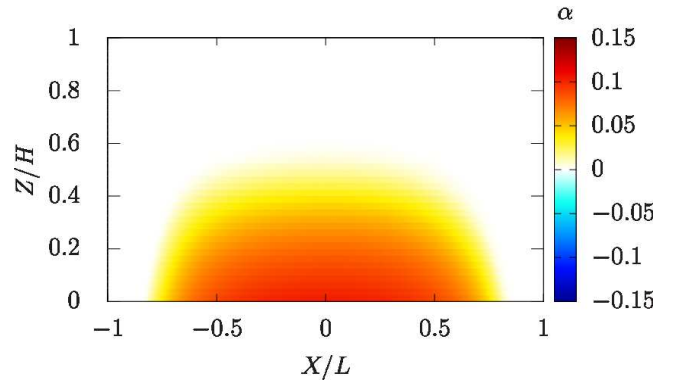}
    \caption{$\alpha$ as a function of $(X,Z)$ after the external oscillation with $N_c = 1/2$.}
    \label{fig:a_1/2}
\end{figure}

Figure \ref{fig:a_1/2} displays $\alpha$ as a function of $(X,Z)$ after the external oscillation with $N_c = 1/2$.
$\alpha$ is positive near the center of the bottom because the equivalent tensile stress $\sigma_{\rm e}$ at the center of the bottom maximizes and exceeds the yield stress $\sigma_Y$ under the horizontal force owing to the external oscillation.
The symmetric $\alpha$ in Eq. \eqref{eq:sN} explains the nearly symmetric distribution of $s_{xx}$ near the bottom as shown in Fig. \ref{fig:s_1/2}.

\begin{figure}[htbp]
	\includegraphics[width=1.0\linewidth]{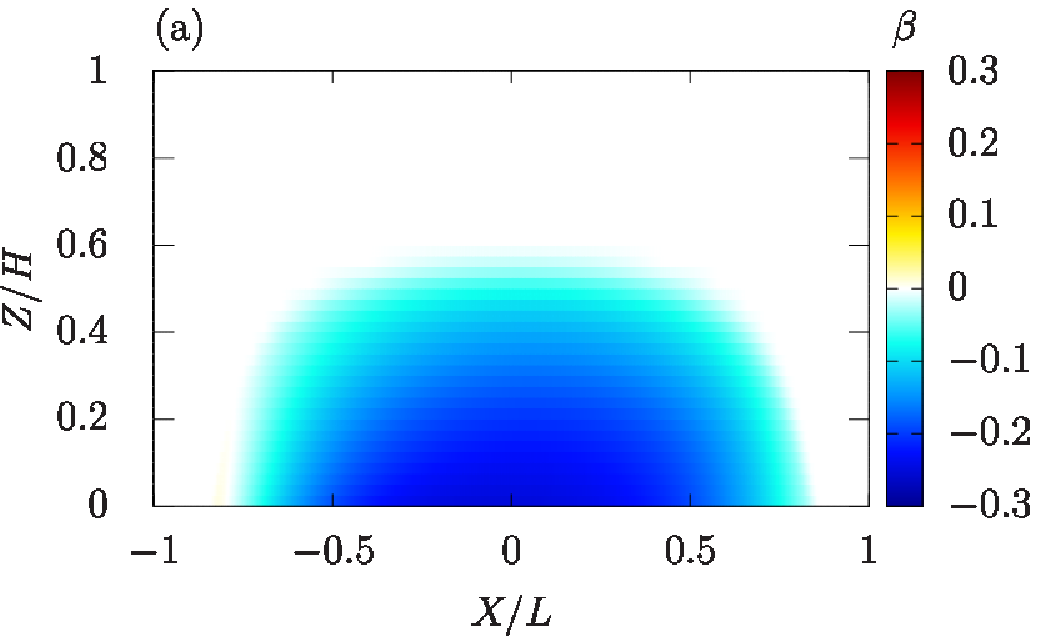}
	\includegraphics[width=1.0\linewidth]{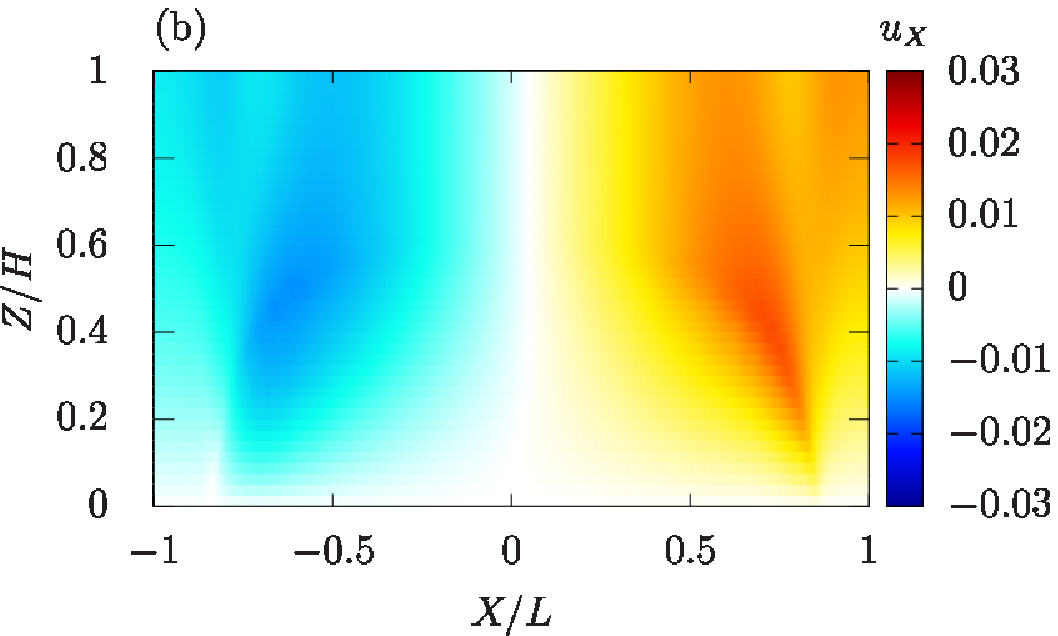}
    \caption{
      (a) $\beta$ as a function of $(X,Z)$ after the external oscillation with $N_c = 1/2$.
      (b)   $u_X$ as a function of $(X,Z)$ after the external oscillation with $N_c = 1/2$.}
    \label{fig:b_uX_0.5}
\end{figure}

In Fig. \ref{fig:b_uX_0.5} (a), we show $\beta$ as a function of $(X,Z)$ after the external oscillation for $N_c = 1/2$.
For $N_c = 1/2$, the external force is applied in the negative $X$-axis direction before the external oscillation is stopped, which leads to the negative shear strain $\partial_Z u < 0$ getting associated with the plastic shear strain $\beta<0$ near the center of the bottom, as shown in Fig. \ref{fig:b_uX_0.5} (a).
The plastic shear strain $\beta$ causes negative displacement $u<0$ in the $X$-axis direction except for the boundaries, which results in the asymmetric $u_X$ as shown in \ref{fig:b_uX_0.5} (b).
This asymmetric $u_X$ in Eq. \eqref{eq:sL} explains the asymmetric residual stress near the surface in Fig. \ref{fig:s_1/2}.

\subsection{Residual stress for $N_c = 2$}
\label{sec:two}

\begin{figure}[htbp]
	\includegraphics[width=1.0\linewidth]{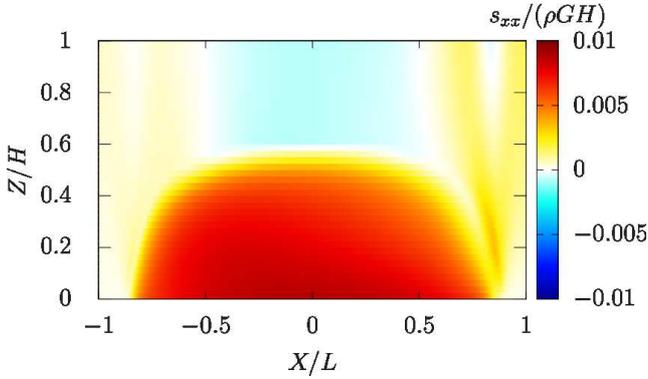}
    \caption{$s_{xx}$ as a function of $(X,Z)$ after the external oscillation with $N_c = 2$.}
    \label{fig:s_2}
\end{figure}

In Fig. \ref{fig:s_2}, we plot $s_{xx}$ as a function of $(X,Z)$ after the external oscillation with $N_c = 2$ (i.e. two cycles).
$s_{xx}$ near the bottom for $N_c = 2$ is higher than that for  $N_c = 1/2$.
The distribution of $s_{xx}$ becomes almost symmetric even near the surface, which is consistent with the symmetric crack patterns shown in experiments with sufficient cycles of the external oscillation \cite{Nakahara18}.

\begin{figure}[htbp]
	\includegraphics[width=1.0\linewidth]{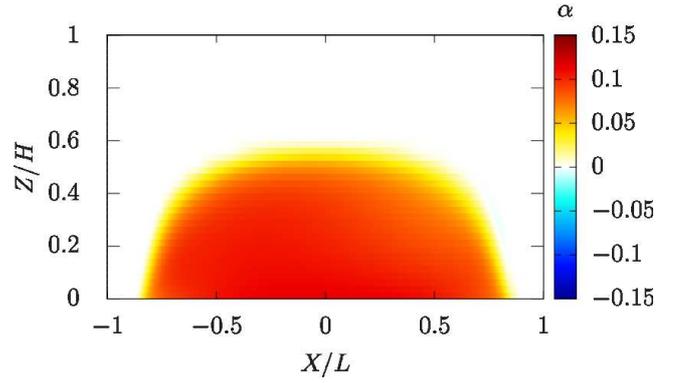}
    \caption{$\alpha$ as a function of $(X,Z)$ after the external oscillation with $N_c = 2$.}
    \label{fig:a_2}
\end{figure}

Figure \ref{fig:a_2} displays $\alpha$ as a function of $(X,Z)$ after the oscillation with $N_c = 2$.
As shown in \ref{sec:steady}, $\alpha$ monotonically increases under shear.
Therefore, $\alpha$ for $N_c = 2$ becomes larger than that for $N_c = 1/2$, which leads to an increase in $s_{xx}$ near the bottom, as shown in Fig. \ref{fig:s_2}.

\begin{figure}[htbp]
	\includegraphics[width=1.0\linewidth]{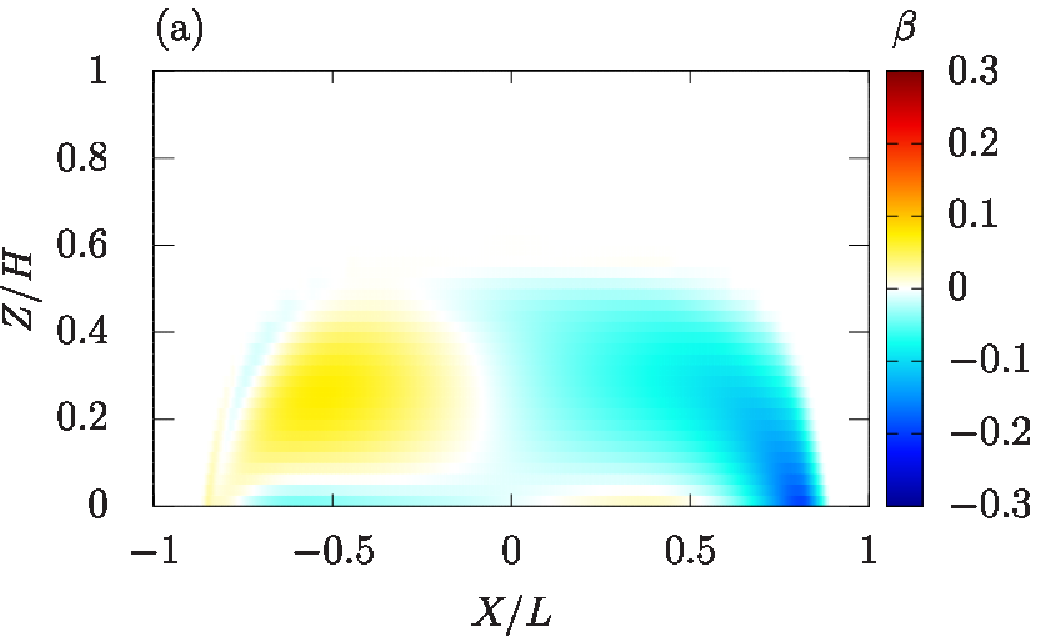}
	\includegraphics[width=1.0\linewidth]{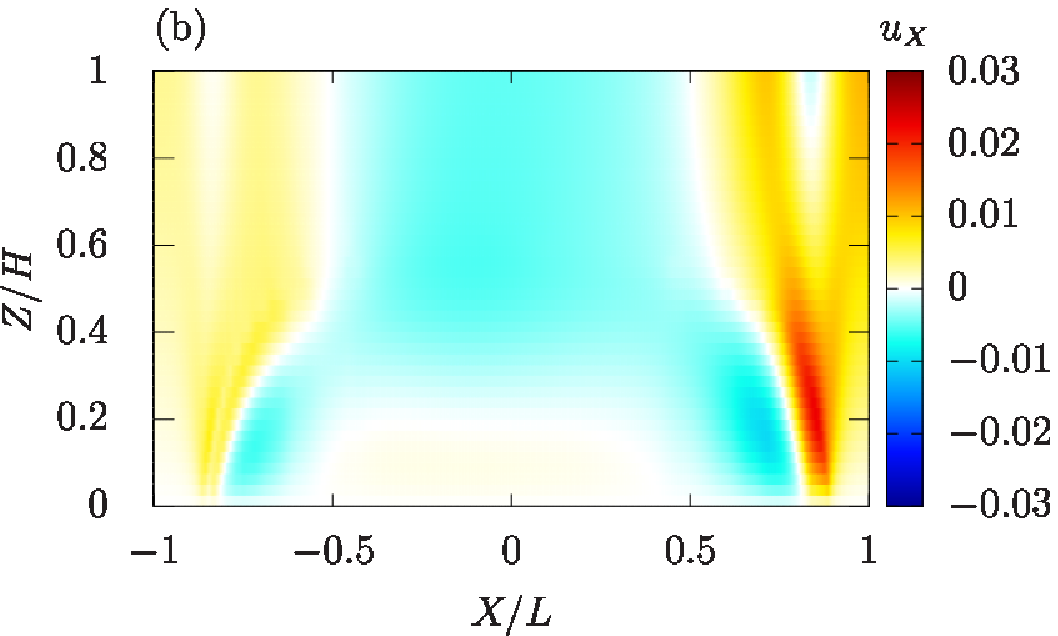}
    \caption{
      (a) $\beta$ as a function of $(X,Z)$ after the external oscillation with $N_c = 2$.
      (b)   $u_X$ as a function of $(X,Z)$ after the external oscillation with $N_c = 2$.}
    \label{fig:b_uX_2}
\end{figure}

In Fig. \ref{fig:b_uX_2} (a), we plot $\beta$ for $N_c = 2$ as a function of $(X,Z)$.
For $N_c = 2$, the plastic deformation is accumulated during the oscillation and the distribution of $\beta$ becomes nearly antisymmetric, which leads to the symmetric $u_X$ as shown in \ref{fig:b_uX_0.5} (b).
This symmetric $u_X$ explains the symmetric stress distribution near the surface, as shown in Fig. \ref{fig:s_2}.

\subsection{Dependence of asymmetry on $N_c$}
\label{sec:number}

\begin{figure}[htbp]
  \centering
  \includegraphics[width=0.9\linewidth]{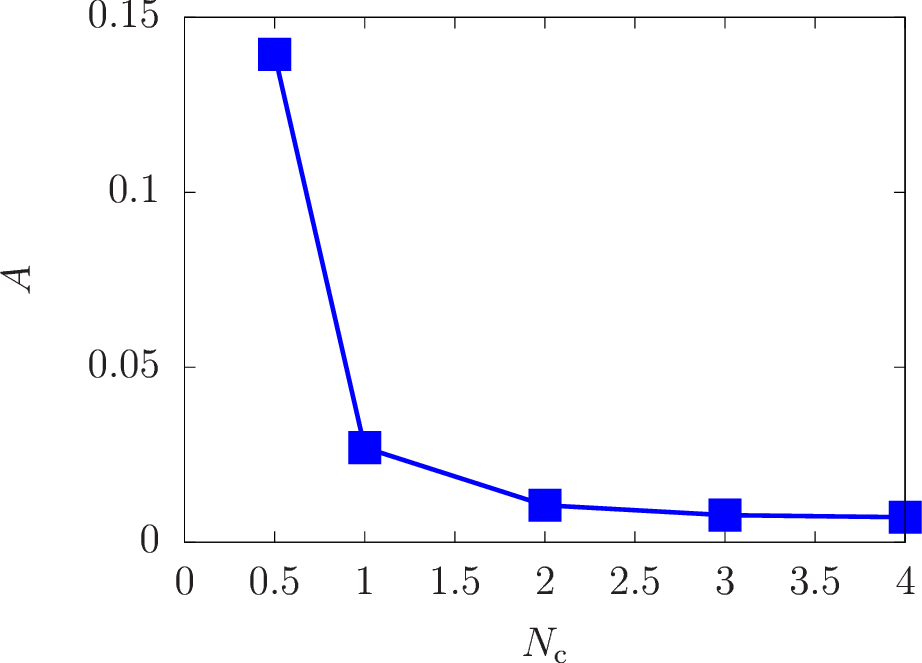}
  \caption{$A$ as a function of $N_c$ after the external oscillation.}
  \label{fig:A}
\end{figure}

The asymmetric part of the deviatoric stress is given by
\begin{eqnarray}
  s_{xx}^{\rm (a)}(X,Z) = \left \{ s_{xx}(X,Z)-s_{xx}(-X,Z) \right \}/2. 
\end{eqnarray}
Here, we define a parameter
\begin{equation}
A =  \frac{
  \int_{-L}^{L} dX \int_0^{H} dZ |s_{xx}^{\rm (a)}|^2
}
{
  \int_{-L}^{L} dX \int_0^{H} dZ |s_{xx}|^2
}
\end{equation}
characterizing the asymmetry of the stress distribution.
In Fig. \ref{fig:A}, we show $A$ after the external oscillation as a function of $N_c$.
$A$ decreases with increasing $N_c$, which is consistent with the symmetry change of the crack patterns in the experiments \cite{Nakahara18}.
Figure \ref{fig:A} indicates that the nonlinear effect becomes dominant
for $N_c \ge 1$.
A similar dependence is obtained for different values of $\mu$.
The sign of the residual stress in the quasi-linear analysis is reversed when the direction of the external force is reversed \cite{Otsuki}, while the stress field is independent of the direction of the external force in the nonlinear analysis \cite{Ooshida08,Ooshida09}.
Hence, we consider that the quasi-linear effect is gradually canceled when the external force is applied in both directions for $N_c \ge 1$, which leads to the fast decrease of $A$ in Fig. \ref{fig:A}.

\section{Summary and discussions}
\label{sec:discussion}

In this study, we numerically investigated the residual stress of a paste after an external oscillation based on an elastoplastic model. 
The residual stress remains as a memory of the oscillation, which leads to crack patterns perpendicular to its direction \cite{Nakahara05,Nakahara06a}.
The residual stress distribution is asymmetric when the number of cycles in the oscillation is small. 
The symmetry of the residual stress is enhanced by increasing the number of cycles, which is consistent with the results of the experiments in Ref. \cite{Nakahara18}.

The plastic deformation induced by the oscillation remains until the formation of cracks \cite{Goehring}.
The desiccation after the oscillation enhances the residual stress caused by the plastic deformation as shown in Ref. \cite{Kitsunezaki16,Kitsunezaki17}.
Thus, we expect that the dependence of the stress symmetry on the number of cycles can be detected in an experiment using the method in Ref. \cite{Kitsunezaki16,Kitsunezaki17}.
Such an experiment will verify the validity of our theory.

Different crack patterns appear depending on the types of powders and ways of applying external forces \cite{Nakahara06b,Matsuo,Nakayama,Akiba}, which is not explained by the present theory.
In our analysis, the process of crack formation and the displacement in the $y$ direction are neglected because we have restricted our attention to the residual stress leading to the perpendicular crack when an external force is applied in one direction.
An extension of the models of drying crack patterns \cite{Kitsunezaki,Ito14a,Ito14b,Halasz} incorporating the effect of the plastic deformation discussed in this study will help understand these crack patterns.

Recent experiments using micro-focus X-ray computerized tomography have revealed that the configuration of microscopic particles in a paste becomes anisotropic after the external oscillation \cite{Kitsunezaki17a}.
This anisotropy is considered as a microscopic memory effect.
The relationship between the microscopic memory effect and the macroscopic residual stress discussed in this study is unclear.
An extension of microscopic theories such as the mode-coupling theory \cite{Miyazaki,Fuchs,Ballauff,Mohan,Fritschi,Mohan17,Moghimi} and the pair distribution function theory \cite{Otsuki06}, predicting the macroscopic constitutive relation, may clarify this relationship.

\begin{acknowledgements}
The authors thank A. Nakahara, S. Kitsunezaki, R. Tarumi, H. Hayakawa, and T. Ooshida for fruitful discussions. 
This work was supported by JSPS KAKENHI Grant Numbers JP16H04025, JP19K03670, and JP21H01006.
\end{acknowledgements}

\section*{Author contribution statement}
J.M. carried out the simulations. Both authors developed the theory and wrote the manuscript.
\appendix

\section{Left Cauchy--Green Tensor}
\label{sec:CGT}

In this Appendix, we derive Eq. \eqref{LCG4} using the method used in Ref. \cite{Ooshida08}.
We first represent the natural metric tensor $g_{ij}^\natural$ as a dot product of orthogonal vectors.
$\bm{g}^\natural$, satisfying Eq. \eqref{Natural metric 1}, is a positive definite symmetric matrix with positive eigenvalues $\lambda_1^2$, $\lambda_2^2$, and $\lambda_3^2$, whose corresponding eigenvectors are $\bm{q}_1$, $\bm{q}_2$, and $\bm{q}_3$, respectively.
A matrix $\mathbf{Q} = [\bm{q}_1 \ \bm{q}_2 \ \bm{q}_3]$ is given by
\begin{eqnarray}
\bm{g}^\natural \mathbf{Q} =
 \mathbf{Q} 
\left[ 
\begin{array}{ccc}
\lambda_1^2 & 0 & 0 \\
0 & \lambda_2^2 & 0 \\
0 & 0 & \lambda_3^2 \\
\end{array} 
\right]
\end{eqnarray}
and
\begin{eqnarray}
\mathbf{Q}^{-1} = \mathbf{Q}^T,
\end{eqnarray}
which indicates
\begin{eqnarray}
\bm{g}^\natural =
 \mathbf{Q} 
\left[ 
\begin{array}{ccc}
\lambda_1^2 & 0 & 0 \\
0 & \lambda_2^2 & 0 \\
0 & 0 & \lambda_3^2 \\
\end{array} 
\right] \mathbf{Q}^T.
\end{eqnarray}
It should be noted that $\mathbf{Q}$ is an orthogonal matrix.
Here, we define a symmetric matrix
\begin{eqnarray}
 \mathbf{P} =
 \mathbf{Q}
\left[ 
\begin{array}{ccc}
\lambda_1 & 0 & 0 \\
0 & \lambda_2 & 0 \\
0 & 0 & \lambda_3 \\
\end{array} 
\right] \mathbf{Q}^T.
\end{eqnarray}
The natural metric tensor $\bm{g}^\natural$ satisfies
\begin{eqnarray}
\bm{g}^\natural = \mathbf{P} \mathbf{P}.
\label{P}
\end{eqnarray}
Here, we introduce independent vectors $\bm{p}_1$, $\bm{p}_2$, and $\bm{p}_2$ satisfying
\begin{eqnarray}
\mathbf{P}  = [\bm{p}_1 \ \bm{p}_2 \ \bm{p}_3].
\end{eqnarray}
The component of $\bm{g}^\natural$ is given by
\begin{eqnarray}
\label{gijp}
g_{ij}^\natural =\bm{p}_i \cdot \bm{p}_j.
\end{eqnarray}

From Eqs. \eqref{Natural metric 1} and \eqref{gijp}, we obtain
\begin{eqnarray}
d\bm{r}^\natural  = \bm{p}_i dX^i.
  \label{dX}
\end{eqnarray}
Because $\bm{p}_1$, $\bm{p}_2$, and $\bm{p}_3$ are independent, there exist dual vectors $\bm{p}_*^1$, $\bm{p}_*^2$, and $\bm{p}_*^3$ satisfying $\bm{p}_i \cdot \bm{p}_*^j  = \delta_i^j$.
It should be noted that the inverse of the component matrix of the metric tensor $(g_{ij}^{\natural} )$ satisfying Eq. \eqref{Natural metric 2} is represented by
\begin{eqnarray}
\label{gijip}
g^{ij}_\natural =\bm{p}_*^i \cdot \bm{p}_*^j.
\end{eqnarray}
Taking a dot product of Eq. \eqref{dX} and $\bm{p}_*^i$, we find
\begin{eqnarray}
dX^i =  \bm{p}_*^i  \cdot d\bm{r}^\natural.
\end{eqnarray}
Substituting this equation into Eq. \eqref{dr}, we obtain
\begin{eqnarray}
d\bm{r} = 
 ( \partial _{i} \bm{r} ) \bm{p}_*^i  \cdot d\bm{r}^\natural
= \left \{( \partial _{i} \bm{r} )  \otimes 
\bm{p}_*^i \right \} d\bm{r}^\natural
\label{drdrn}
\end{eqnarray}
From Eqs. \eqref{Deformation gradient} and \eqref{drdrn}, we find
\begin{align}
\label{Deformation gradient:dr}
\mathbf{F} = ( \partial _{i} \bm{r} )  \otimes \bm{p}_*^i.
\end{align}
Substituting this equation into Eq. \eqref{B} with Eq. \eqref{gijip}, we obtain Eq.  \eqref{LCG4}.

\section{Stretching Tensor}
\label{sec:ST}

In this Appendix, we derive Eq. \eqref{strain rate}.
First, we introduce 
\begin{eqnarray}
  \nabla X^i = \frac{\partial X^i}{ \partial \bm{r}},
\end{eqnarray}
which satisfies
\begin{eqnarray}
(\partial_i \bm{r}) \cdot \nabla X^j = \delta_i^j
\label{dual 1}
\end{eqnarray}
and
\begin{eqnarray}
(\partial_i \bm{r}) \otimes \nabla X^i = \mathbf{I}.
\label{dual 2}
\end{eqnarray}
A component of the inverse matrix of $\boldsymbol{g}$ satisfying Eq. \eqref{Orthogonal} is given by the dot product of $\left \{ \nabla X^i \right \}$:
\begin{eqnarray}
  g^{ij} = \nabla X^i \cdot \nabla X^j.
  \label{inverse natural metric}
\end{eqnarray}
Using Eq. \eqref{dual 2} with Eq. \eqref{inverse natural metric}, we obtain
\begin{eqnarray}
  \nabla X^i & = & \mathbf{I} \nabla X^i  \nonumber \\
& = & \left \{ (\partial_j \bm{r}) \otimes \nabla X^j \right \} \nabla X^i, 
\nonumber \\
& = &\left \{ \nabla X^i \cdot \nabla X^j \right \} (\partial_j \bm{r}), 
\nonumber \\
& = & g^{ij} \partial_j \bm{r}.
\label{nabla X}
\end{eqnarray}
Substituting Eq. \eqref{nabla X} into Eq. \eqref{dual 2},
\begin{eqnarray}
  g^{ij} (\partial_i \bm{r}) \otimes (\partial_j \bm{r})  = \mathbf{I}.
  \label{gr}
\end{eqnarray}

Taking the dot product of Eq. \eqref{dr} with $ \nabla X^i$, we obtain
\begin{eqnarray}
dX^i=\nabla X^i \cdot d\bm{r}.
\end{eqnarray}
Substituting this equation into Eq. \eqref{dv}, we obtain
\begin{eqnarray}
  d\bm{v} = (\partial_i \bm{v})(\nabla X^i \cdot d\bm{r})
  = \left \{ (\partial_i \bm{v}) \otimes \nabla X^i\right \}d\bm{r}.
\end{eqnarray}
Comparing this equation with Eq. \eqref{Velocity gradient}, we find
\begin{eqnarray}
  \boldsymbol{W}  = (\partial_i \bm{v}) \otimes \nabla X^i
  \label{W}
\end{eqnarray}
Substituting Eq. \eqref{W} into Eq. \eqref{stretching} with Eq. \eqref{W}, we obtain
\begin{eqnarray}
  \boldsymbol{D}  
  & = & \frac{1}{2}g^{ij}
  \left \{
  (\partial_i \bm{v}) \otimes (\partial_j \bm{r}) 
  +
  (\partial_i \bm{r}) \otimes (\partial_j \bm{v}) 
\right \}
\nonumber \\
  & = & \frac{1}{2}g^{ij}
\partial_t
  \left \{
  (\partial_i \bm{r}) \otimes (\partial_j \bm{r}) 
\right \}
\label{D}
\end{eqnarray}
Differentiating Eq. \eqref{gr} by $t$, we find
\begin{equation}
  g^{ij} \partial_t \left \{ (\partial_i \bm{r}) \otimes (\partial_j \bm{r}) \right \}
  = -\left ( \partial_t g^{ij}\right ) \left \{ (\partial_i \bm{r}) \otimes (\partial_j \bm{r})  \right \}.
\end{equation}
Substituting this equation into Eq. \eqref{D}, we obtain Eq. \eqref{strain rate}.

\section{Elasto-plastic part of stress tensor}
\label{sec:stress}

In this appendix, we derive Eq. \eqref{stress 2}.
Substituting Eq. \eqref{Strain Energies} into Eq. \eqref{Constitutive}, we obtain
\begin{align}
\label{Constitutive 2}
\boldsymbol{\sigma}^{\rm (EP)}  = 
\frac{\Sigma}{J } 
\frac{\partial J}{\partial \tilde{\mathbf{F}}}\  \tilde{\mathbf{F}}^T
+
\frac{\partial \Sigma}{\partial \tilde{\mathbf{F}}}\  \tilde{\mathbf{F}}^T.
\end{align}
Using a formula 
\begin{align}
\label{Det A}
  \frac{\partial \det {\mathbf{A}}}{\partial \mathbf{A}}\  {\mathbf{A}}^T
  = \left ( \det {\mathbf{A}} \right ) \mathbf{I} 
\end{align}
for a tensor $\mathbf {A}$,
$J = \det \tilde{\mathbf{F}}$ satisfies
\begin{align}
\frac{\Sigma}{J } 
\frac{\partial J}{\partial \tilde{\mathbf{F}}}\  \tilde{\mathbf{F}}^T
=
\Sigma \mathbf{I}.
\label{First}
\end{align}

The strain energy per unit volume in the current coordinates $\Sigma$ is a function of the rotational invariants $I_1$ and $I_3$ of the left Cauchy--Green strain tensor $\mathbf{B}$.
Hence, the Cartesian component of the second term in Eq. \eqref{Constitutive 2} satisfies
\begin{align}
\left ( \frac{\partial \Sigma}{\partial \tilde{\mathbf{F}}}\  \tilde{\mathbf{F}}^T
\right )_{ij}
=
\left \{
\frac{\partial \Sigma}{\partial I_1} 
\frac{\partial I_1}{\partial B_{lm}} 
+
\frac{\partial \Sigma}{\partial I_3}
\frac{\partial I_3}{\partial B_{lm}} 
\right \}
\frac{\partial B_{lm}}{\partial \tilde{F}_{ik}}\  \tilde{F}_{jk}.
\label{W F}
\end{align}
Here, we obtain 
\begin{align}
\frac{\partial I_1}{\partial  \mathbf{B}}
= \mathbf{I}
  \label{I1B}
\end{align}
and
\begin{align}
  \frac{\partial I_3}{\partial \mathbf{B}}\  
  = I_3 {\mathbf{B}}^{-1}
  \label{I3B}
\end{align}
from Eq. \eqref{Det A} and $\mathbf{B}^T = \mathbf{B}$.
Using Eqs. \eqref{B}, \eqref{gijip}, and \eqref{Deformation gradient:dr},
we find
\begin{align}
 \mathbf{B}=
  \tilde{\mathbf {F}} {\bm g}_\natural \tilde{\mathbf {F}}^T.
\label{BFP}
\end{align}
From Eq. \eqref{BFP},
we obtain
\begin{align}
\frac{\partial B_{lm}}{\partial \tilde{F}_{ik}}\  \tilde{F}_{jk}
  = \delta_{il}B_{jm} + \delta_{im}B_{lj}.
\label{BFF}
\end{align}
Substituting Eqs. \eqref{I1B}, \eqref{I3B}, and \eqref{BFF} into
Eq. \eqref{W F}, we derive
\begin{align}
\frac{\partial \Sigma}{\partial \tilde{\mathbf{F}}}
\tilde{\mathbf {F}}^T
=
2   
\left \{
\frac{\partial \Sigma}{\partial I_1}
\mathbf{B}
+I_3 
\frac{\partial \Sigma}{\partial I_3}\mathbf{I}
  \right \}.
\label{Second}
\end{align}
Substituting Eqs. \eqref{First} and \eqref{Second} into Eq. \eqref{Constitutive 2}, we obtain
\begin{align}
\label{Constitutive 3}
\boldsymbol{\sigma}^{\rm (EP)}  = \Sigma \mathbf{I}  +
2   
\left \{
\frac{\partial \Sigma}{\partial I_1}
\mathbf{B}
+I_3 
\frac{\partial \Sigma}{\partial I_3}\mathbf{I}
  \right \}.
\end{align}
Substituting Eq. \eqref{Hadamard 1} into this equation, we derive Eq. \eqref{stress 2}.

\section{Stress under uniform steady shear}
\label{sec:steady}

In this appendix, we discuss the rheological properties described by Eqs.  \eqref{stress 2} and \eqref{Maxwell 5} with Eq. \eqref{LCG4} under simple steady shear:
\begin{eqnarray}
\label{Simple shear}
\bm{r} ( X , Y,  Z, t ) = 
\left[ 
\begin{array}{ccc}
  X + \gamma(t) Z \\
Y \\
Z \\
\end{array} 
\right]_{\rm C},
\end{eqnarray}
where the shear strain is given by
\begin{eqnarray}
\label{Shear rate}
\gamma(t) = \dot \gamma t
\end{eqnarray}
with the shear rate $\dot \gamma$.
The natural metric tensor $\mathbf{g}^{\natural}$ is represented by Eq. \eqref{Natural metric3} with $\alpha = \beta=0$ at $t=0$.

\begin{figure}[htbp]
  \centering
  \includegraphics[width=1.0\linewidth]{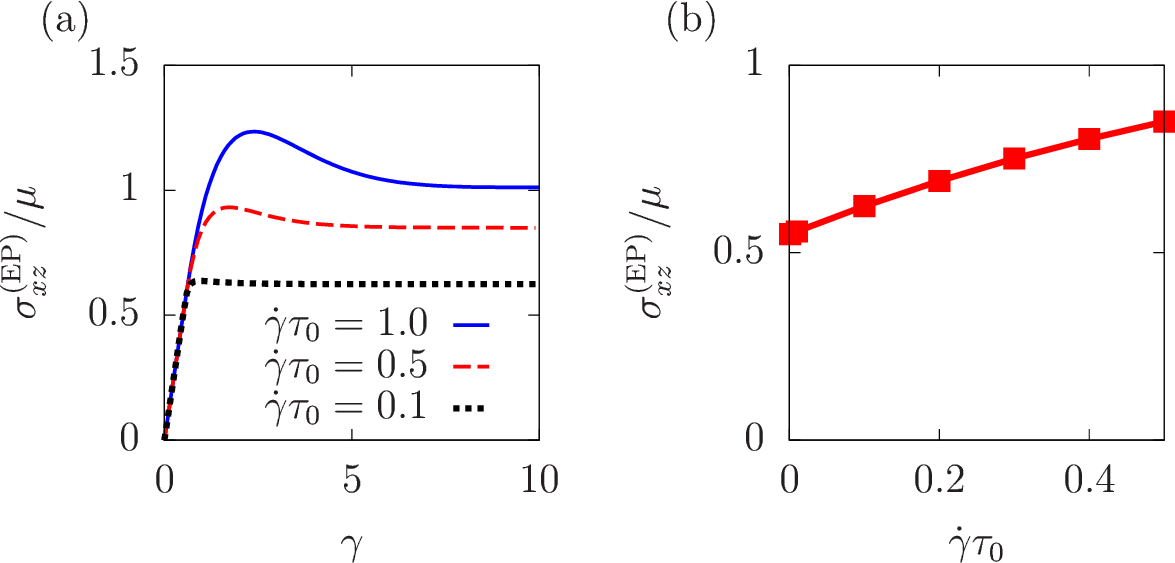}
  \caption{
    (a) $\sigma_{xz}^{\rm (EP)}$
  against the shear strain $\gamma$
  for $\sigma_{\rm Y}/\mu=1.0$ with different $\dot \gamma \tau_0$.
    (b) $\sigma^{\rm (EP)}_{xz}$
  against the shear strain rate $\dot \gamma$
  for $\sigma_{\rm Y}/\mu=1.0$ with different $\dot \gamma \tau_0$ in the steady state.
    }
  \label{fig:s_ga}
\end{figure}
Substituting Eqs. \eqref{Simple shear} and \eqref{Natural metric4} into Eqs. \eqref{stress 2} and \eqref{Maxwell 5} with Eq. \eqref{LCG4}, we obtain the time evolution equations for the shear stress $\sigma_{xz}^{\rm (EP)}$ as
\begin{eqnarray}
  \partial_t \sigma_{xz}^{\rm (EP)} & = & \mu e^{-\alpha} \dot \gamma - 
  \tau^{-1}(\sigma_{\rm e}) \sigma_{xz}^{\rm (EP)}  \\
  \partial_t \alpha & = & \tau^{-1}(\sigma_{\rm e}) \left (  1 - \frac{3}{I_1} e^\alpha \right ) \label{eq:alpha}\\
  \partial_t \beta & = & \tau^{-1}(\sigma_{\rm e}) \left (  \frac{3}{I_1} \gamma
  - \beta \right ) \label{eq:beta}
\end{eqnarray}
with 
\begin{eqnarray}
  I_1 = 3 + 2 (\cosh \alpha - 1) + e^\alpha \left ( \sigma_{xz}^{\rm (EP)} / \mu\right )^2.
\end{eqnarray}
Here, $\sigma_{\rm e}$ is given by 
\begin{eqnarray}
  \label{Equivalent shear}
  \sigma_{\rm e} =  \sqrt{3 \left ( \sigma_{xz}^{\rm (EP)}\right ) ^2 + \Xi^2}
\end{eqnarray}
with
\begin{eqnarray}
  \Xi^2 & = & \frac{\mu^2  e^{2\alpha}}{2} \left \{ 1 + \left ( \frac{\sigma_{xz}^{\rm (EP)}}{\mu} \right )^2 - e^{-\alpha} \right \}^2 \nonumber \\
  & & + \frac{\mu^2}{2} (e^{-\alpha}-1)^2 \nonumber \\
  & &  + \frac{\mu^2}{2} \left \{ e^\alpha+e^\alpha \left ( \frac{\sigma_{xz}^{\rm (EP)}}{\mu}  \right )^2
  - e^{-\alpha}\right \}^2.
\end{eqnarray}
It should be noted that
\begin{eqnarray}
  \partial_t \alpha & \simeq & \tau^{-1}(\sigma_{\rm e}) \left \{  -\alpha  + \frac{1}{3} \left (\frac{ \sigma_{xz}^{\rm (EP)}}{ \mu } \right )^2 \right \}
\end{eqnarray}
for $\alpha \ll 1$ and $\sigma_{xz}^{\rm (EP)} \ll  \mu$, which indicates that $\alpha$ monotonically increases under steady shear.

In Fig. \ref{fig:s_ga}(a), we plot the shear stress $\sigma^{\rm (EP)}_{xz}$ against the shear strain $\gamma$ for $\sigma_{\rm Y}/\mu=1.0$ with different $\dot \gamma \tau_0$.
For sufficiently small $\gamma$, $\sigma^{\rm (EP)}_{xz}$ is almost proportional to $\gamma$.
As $\gamma$ increases, $\sigma^{\rm (EP)}_{xz}$ exhibits a peak and converges to a steady state.
$\sigma^{\rm (EP)}_{xz}$ and $\gamma$ at the peak increase with increasing $\dot \gamma$.
This behavior is qualitatively similar to the stress--strain curve in experiments \cite{Aken} and numerical simulations \cite{Varnik}.

Figure \ref{fig:s_ga}(b) displays the shear stress $\sigma_{xz}^{\rm (EP)}$ against the shear strain rate $\dot \gamma$ for $\sigma_{\rm Y}/\mu=1.0$ in the steady state.
$\sigma^{\rm (EP)}_{xz}$ monotonically increases and obeys $\sigma_{xz}^{\rm (EP)} = \frac{\sigma_{\rm Y}}{\sqrt{3}} +  \mu\dot \gamma \tau_0$, which is consistent with the behavior of a Bingham plastic fluid \cite{Bingham}.



\end{document}